\documentclass[jkps,preprint,fleqn,showpacs,showkeys,10pt,twocolumn]{revtex4}

\usepackage[pdftex]{graphicx}
\usepackage{amssymb}
\usepackage{amsmath}
\usepackage{bm}

\usepackage{color,hyperref}

\begin{document}
\setcounter{page}{1}
\title[]{Orbit Error Correction on the High Energy Beam Transport Line at the KHIMA Accelerator System}
\author{Chawon \surname{Park}}
\email[E-mail: ]{parknkim@kirams.re.kr}
\author{Heejoong \surname{Yim}}
\author{Garam \surname{Hahn}}
\author{Dong Hyun \surname{An}}
\email[E-mail: ]{ectroan@kirams.re.kr}
\affiliation{Division of Heavy Ion Accelerator, Korea Institute of Radiological \& Medical Sciences, Seoul
139-706}
\begin{abstract}
For the purpose of treatment of various cancer and medical research, the synchrotron based medical machine under the Korea Heavy Ion Medical Accelerator (KHIMA) project have been conducted and is going to treat the patient at the beginning of 2018. The KHIMA synchrotron is designed to accelerate and extract the carbon ion (proton) beam with various energy range, 110 up to 430 MeV/u (60 up to 230 MeV).
A lattice design and beam optics studies for the High Energy Beam Transport (HEBT) line at the KHIMA accelerator system have been carried out with WinAgile and the MAD-X codes.  Because the magnetic field errors and the mis-alignments introduce to the deviations from the design parameters, these error sources should be treated explicitly and the sensitivity of the machine's lattice to different individual error sources is considered. Various types of errors which are static and dynamic one have been taken into account and have been consequentially corrected with a dedicated correction algorithm by using the MAD-X program. As a result, the tolerances for the diverse error contributions have been specified for the dedicated lattice components in the whole HEBT lines. 
\end{abstract}


\maketitle


\section{INTRODUCTION}
\begin{figure}[h]
\begin{center}
\includegraphics[width=8.0cm]{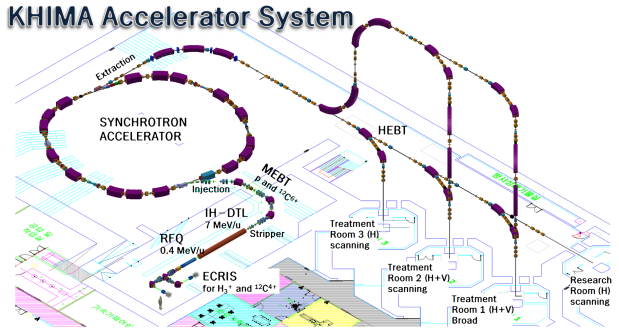}
\caption{(Color online) Schematic layout of KHIMA synchrotron system including each treatment and research rooms.}
\label{fig0}
\end{center}
\end{figure}

The Korea Heavy Ion Medical Accelerator (KHIMA) project at the Korean Institute of Radiological And Medical Sciences (KIRAMS) 
has carried out the development of an accelerator based on synchrotron with multi-ion sources 
for various cancer treatement. The designed synchrotron accelerates the proton beam (the carbon ion, $^{12}C^{6+}$, beam) 
from 60 MeV (110 MeV/$u$) to 230 MeV (430 MeV/$u$). Those energy ranges correspond to the penetration depth of 3.0 cm to 31.0 cm in water. 
A schematic layout of the accelerator center is shown in Fig.~\ref{fig0}. 
At the Electron Cyclotron Resounce Ion Source (ECRIS), ions with a charge to mass ratio q/m = 1/3, either $H^{+}_{3}$ or $^{12}C^{4+}$, 
are generated up to 8.0 keV/$u$. 
These ions are accelerated up to 7 MeV/$u$ through Radio Frequency Quadrupole (RFQ) linac and Interdigital H-mode Drift Tube Linac (IH-DTL). 
At the beginning of Medium Energy Beam Transport (MEBT) two corresponding ions are stripped and fully ionized to either proton or $^{12}C^{6+}$, 
then are transported to the synchrotron being accelerated up to designed energies. 
Each ion is injected into the synchrotron through a multi-turn injection mechanism, accelerated by switching the RF system 
and then extracted into High Energy Beam Transport (HEBT) line by slow resonance extraction scheme~\cite{Extract,HJYim}.

\section{Beam characteristic on the HEBT}
In each HEBT line, the beam is defined by the medical specifications and thus there is a definite range of vertical Twiss functions, 
$2.0 < \beta_{y} < 27 ~\rm{m}, ~\alpha_{y}=0$. For adjusting the beam size in the vertical plane, 
the straightforward method is to be used. The beam size is determined by the traditional geometrical emittance of
the beam originated from synchrotron 
and the beta function at the iso-center. The vertical beam size is calculated as $y = \sqrt{\epsilon_{y} \times \beta_{y}}$~\cite{PIMMS}.
On the other hand the peculiar distribution in the horizontal plane is appeared so called 'bar of charge' in phase space,
because the extracted beam is the segment of the extraction separatrix that could be represented 
as the diameter of an unfilled ellipse. 
Thus by varying the phase advance ($\Delta \mu_x$), namely by changing the orientation of the charge bar 
and consequently its projection to the x-axis determines the horizontal beam size~\cite{PIMMS,MBenedikt,SRossi}. 

\section{Beam optics simulation}
The design concept of KHIMA extraction, HEBT, lines is based on the Proton-Ion Medical Machine Study (PIMMS)~\cite{PIMMS}. 
Three medical treatment rooms and one research oriented irradiation room are prepared for the center. 
The HEBT lines compose the 6 different transport branches with 4 horizontal- and 2 vertical-lines as shown in Fig~\ref{fig0}. 
A slowly extracting beam through electrostatic septum (ES) in the synchrotron ring 
is to be selectively transported into each treatment room.
An integrated system is designed with telescope modules with integer $\pi$ phase advances ($\Delta \mu$). 
The HEBT lines are also based on a modular design taking into account the strong asymmetry 
between two transverse beams. The trapezoidal distribution of the horizontal beam is considered. 
For horizontal beam in phase space, the bar of charge is applied to create an independent control of 
the horizontal beam size by rotating the bar in an unfilled ellipse~\cite{MBenedikt01}. 
 On the other hand, the Gaussian shaped beam distribution is taken into account the vertical beam.

\subsection{Electrostatic Septum to Matching Section}
On the base of the PIMMS report, 
the dispersion vector at the ES is determined by appearance of the charge bar for different momenta of the particle beam. 
In Table~\ref{table:ES}, 
the data for lowest extraction energy are summarized 
and the dispersion ($Dx$) and its derivative (${Dx}^{\prime}$) values are listed. 
At higher energies, the corresponding values are not so different and not significantly gotten out. 

Furthermore, the $Dx$ and ${Dx}^{\prime}$ from the synchrotron are not representative ones
as for the betatron amplitude function. 
The $Dx$ and ${Dx}^{\prime}$ were calculated as listed in Table~\ref{table:ES}, 
thus it could be utilized universally up to matching section.
In principle any values can be selected as an initial values at the ES,
but the useful one should be selected depending on the dedicated machine. 
Because the dispersion function represents the displacement of the off-momentum particles, 
it is natural to be selected as an initial value corresponding to the distance 
between the center of gravity of the on-momentum particles and the center of gravity of the off-momentum ones 
divided by the relative momentum difference as described and calculated in Table~\ref{table:ES}.
\begin{table*}
\caption{Dispersion of the extracted beam at the entry to the ES.} 
\begin{center}
\begin{ruledtabular}
\begin{tabular}{l|cc|cc}
                    &   On resonance particles &  &  Off resonace particles &\\
                    &   $\Delta p/p = 0$ &  & $\Delta p/p = 0.0010$ &\\
\hline
  & Position(m) & Angle(rad) & Position(m)& Angle(rad) \\
\hline
Inner edge of segment  & 0.050000 & 0.001952 & 0.050000 & 0.002048 \\
Outer edge of segment  & 0.057023 & 0.002164 & 0.054715 & 0.002188 \\
Average radial position/angle  & 0.053115 & 0.002058 & 0.052358 & 0.002118 \\ \hline
Radial shift(m)  & \multicolumn{4}{c}{-0.000757}   \\
Angluar shift(rad)  &  \multicolumn{4}{c}{ 0.00013}  \\
Dispersion, $Dx$(m)  &   \multicolumn{4}{c}{0.7407}  \\
Derivative of dispersion, ${Dx}^{\prime}$(m)  & \multicolumn{4}{c}{0.1272}  \\ 
\end{tabular}
\end{ruledtabular}
\end{center}
\label{table:ES}
\end{table*}
\begin{figure}[h]
\begin{center}
\includegraphics[width=7.0cm]{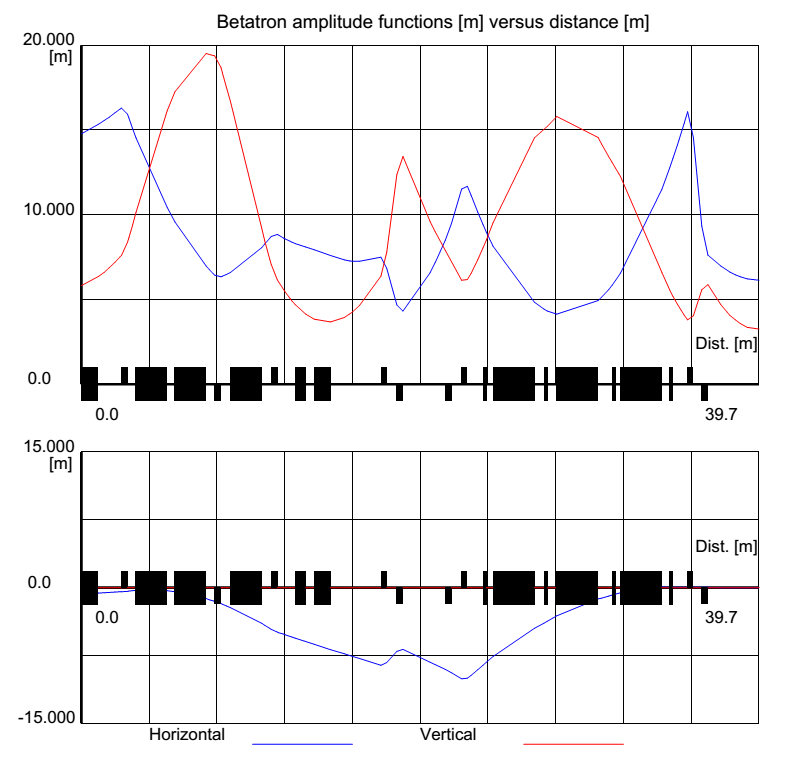}
\caption{(Color online) Betatron amplitude and dispersion function at the matching section for the horizontal and vertical plane, 
where half height rectangular box corresponds to the quadrupole magnet having their polarity 
while the full height one for the dipole magnet.}
\label{fig1}
\end{center}
\end{figure}
According to the determined values in Table~\ref{table:ES}, 
the boundary conditions at the ES are selected.
Fig.~\ref{fig1} shows the distribution of Twiss parameters at the matching module.

\subsection{Horizontal Beam Lines}
\begin{figure}[h]
\begin{center}
\includegraphics[width=7.0cm]{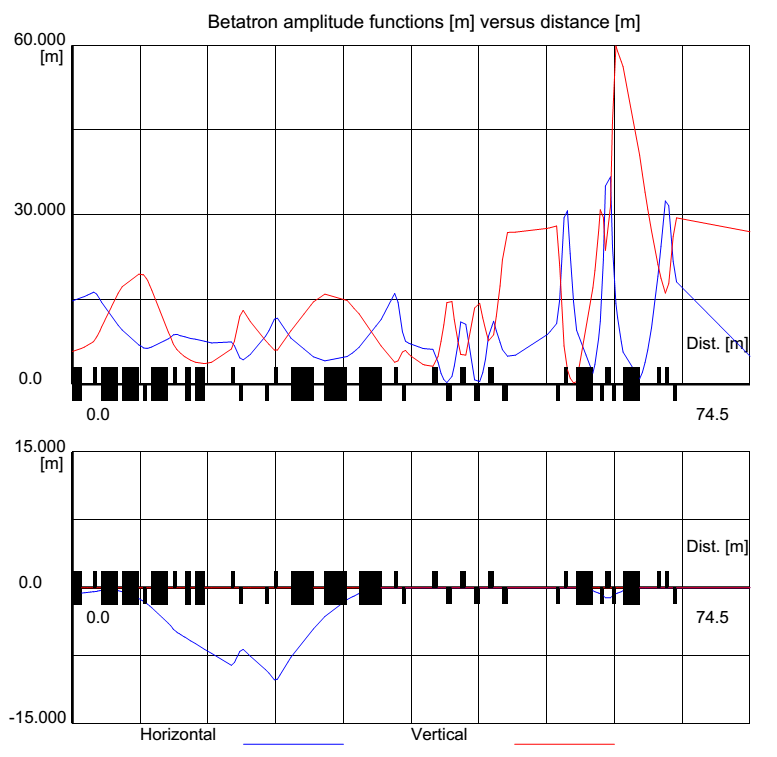}  
\caption{(Color online) Betatron amplitude and the dispersion functions for the 3rd Horizontal (H3) line, 
where $\Delta \mu_{x}=~6.0~\pi$ and $\beta_{y} = 27~m$ at the iso-center are set.}
\label{fig2}
\end{center}
\end{figure}
The common module downstream matching section in the horizontal common line (HC)
is interleaved for controlling beam size at the iso-center.
This module has an important advantage for commissioning and operation. 
As described in previous section,
the vertical beam size at the iso-center is controlled by the vertical $\beta$ function.  
In order to fulfil the requirements,
the module can be tuned to provide the values among $2 < \beta_{y} < 27$ m with $\alpha_{y} = ~0$. 
A minimum of five components are needed to control the five Twiss variables, i.e., 
the horizontal, vertical $\beta$'s and the $\alpha$'s at the exit and the horizontal phase advance. 
As the variable component, each quadrupole magnet in the common line is used.
At least six quadrupoles are needed to facilitate the matching and to have some redundancy. 
This composition of consecutive six quadrupole magnets, which functions as the phase shifter on the horizontal plane 
and also as the stepper on the vertical plane, is named simply as the stepper.

As one of the horizontal lines, 
Fig.~\ref{fig2} shows the Twiss function distributions up to 3rd treatment room through the horizontal H3 line. 

\section{Error analysis}
Because various distortion making elements in the transport line are exist, 
the closed orbit control is a basic constituent for the efficient performance in the beam line. 
A large closed orbit distortion minimizes the available aperture and also affect the dynamics of the beam via non-linear elements. 
Various error sources contribute to the closed orbit distortion.
Most of errors are random in nature, while others are systematic and some may be time or field dependent~\cite{Badano}.  

The results have been calculated by using the accelerator codes, which are the WinAGile~\cite{Agile} or the MAD-X~\cite{Schmidt,Herr}.
Both codes base the correction process on a least square method. 
A series of test runs showed that the results obtained from two program codes are consistent within their statistical uncertainty. 
Thus it is determined to use the MAD-X code representatively in this analysis. 
One of the simulations is performed with a statistics for 1000 randomly generated machines in the vertical transport line (V2).

\subsection{Requirements for Closed Orbit Correction}
The basic guidelines for the closed orbit correction for the HEBT in the KHIMA accelerator are 
that the global closed orbit correction must be within 7.5 mm for both horizontal and vertical planes.
For error analysis in the study, only the static errors are described, 
while the dynamic errors originated from the stability of magnets are small enough to be ignored compared to static ones.
 
\subsection{Alignment and Field Tolerances}
The impact of each error source, so called sensitivity study, was evaluated using the MAD-X simulation code. 
The accepted tolerances for various errors resulting in an orbit distortion are listed in Table~\ref{table:Tolerance}. 
The dipole field error is originated from the packing factor tolerance and the its length error. 
The correctors and position monitors are considered to be affected by several errors. 
The position monitors also include both alignment and reading errors. 
Alignment- and field-errors are applied to all the elements according to Gaussian distribution by cutting 3 standard deviations, 
while the monitor reading errors are given with a uniform distribution as described in Table~\ref{table:Tolerance}.
\begin{table*}
\caption{Accepted tolearances for the magnet elements and position monitors.}
\begin{ruledtabular}
\begin{tabular}{l|c|c}
Error type                    &   Tolerance &  Distribution\\ \hline
Qaudrupole alignment (x, y, z) & 0.3 mm & Truncated Gaussian \\
Qaudrupole tilt & 0.3 mrad & Truncated Gaussian \\
Dipole alignment (x, y, z) & 0.3 mm & Truncated Gaussian \\
Dipole tilt & 0.3 mrad & Truncated Gaussian \\
Integrated dipole field error ($\Delta BL/BL$) & 0.001 & Truncated Gaussian \\
Integrated quadrupole field error  & 0.001 & Truncated Gaussian \\
Monitor reading error & $\pm0.5$ mm & Flat \\
\end{tabular}
\end{ruledtabular}
\label{table:Tolerance}
\end{table*}

\subsection{Monitor and Corrector System}
\begin{figure}[h]
\begin{center}
\includegraphics[width=9.0cm]{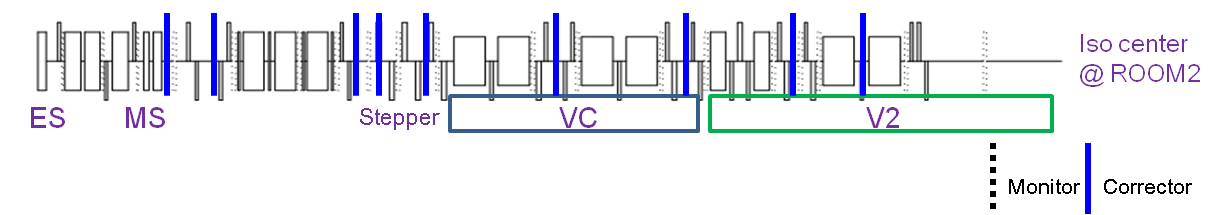}
\caption{(Color online) Beam line layout through V2 with the beam position monitor and corrector.}
\label{fig9}
\end{center}
\end{figure}

The monitor system is correlated with the dual corrector which is the steering activated as a pair in the horizontal and vertical plane. 
The precision and reliability of the measurement system are extremely important, 
because they determine the quality of the corrector together with the precision of the corrector and their maximum magnetic strength.     
The layout through V2 line for optimizing the orbit correction is shown in Fig.~\ref{fig9}.

\subsection{Closed Orbit Estimation and Correction}
\begin{figure}[h]
\begin{center}
\includegraphics[width=9.0cm]{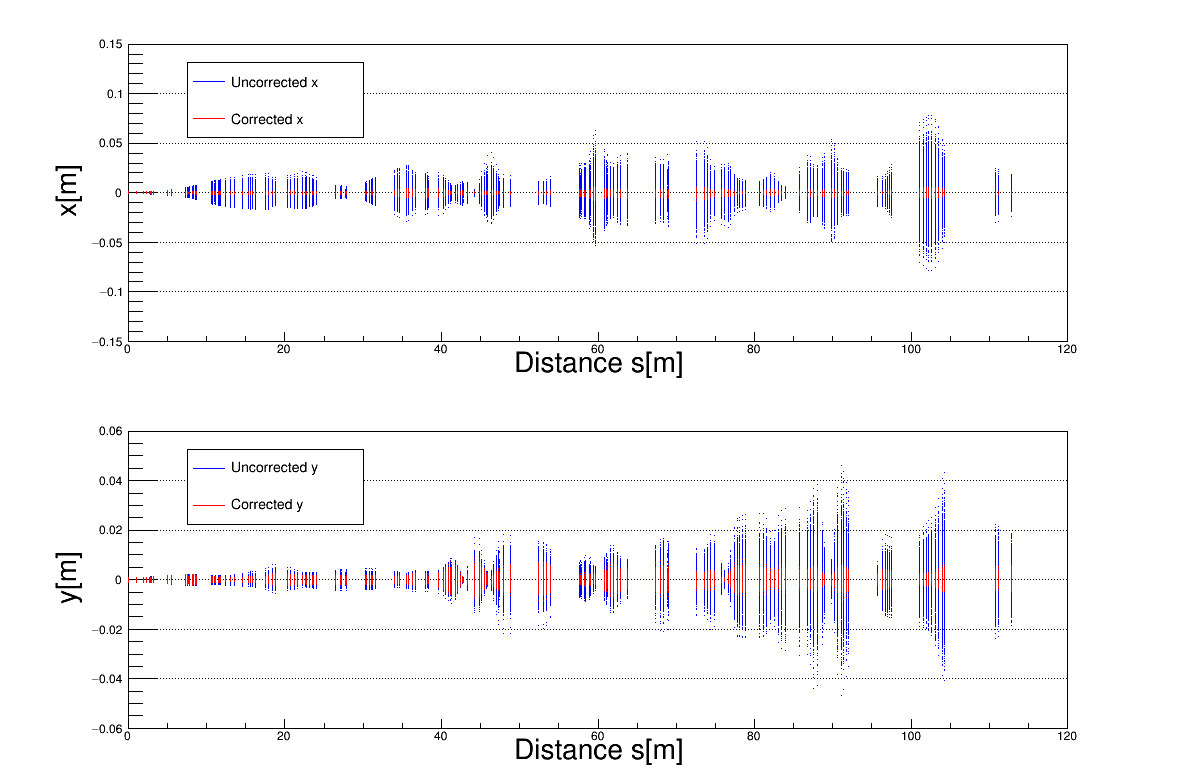}
\caption{(Color online) Excursion distributions through the V2 line before and after correction.}
\label{fig10}
\end{center}
\end{figure}

\begin{figure}[h]
\begin{center}
\includegraphics[width=9.0cm]{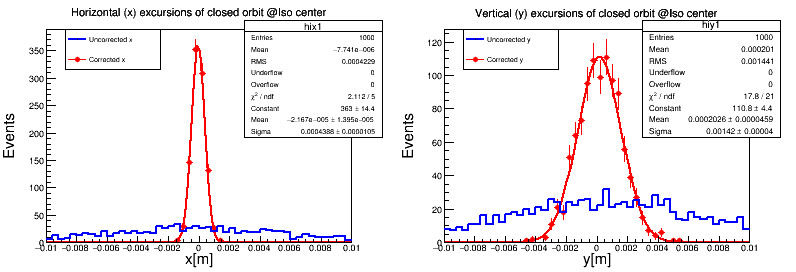}
\caption{(Color online) Excursion distributions of closed orbit at iso-center before and after correction.}
\label{fig11}
\end{center}
\end{figure}

The orbits have been calculated with MAD-X code on the HEBT line through V2.
The mechanism of the transformation is provided by the technique of singular value decomposition (SVD)~\cite{SVD} of matrices.
For testing what various sources of errors have a sensitivity,
the execursed quantity with respect to the sequential beam line is evaluated
before and after correction as shown in Fig.~\ref{fig10}. In the beginning,
the total number of monitors used is 15, while 9 for correctors. 
One thousand of machines with random errors have been analysed before and after the correction on the vertical line through V2.
At iso-center, the beam position is accurate dramatically after correction as shown in Fig.~\ref{fig11}.

The absolute maximum excursions are quoted,
because these values are of more direct interest for the aperture than the peak-to-peak values.
\begin{figure}[h]
\begin{center}
\includegraphics[width=9.0cm]{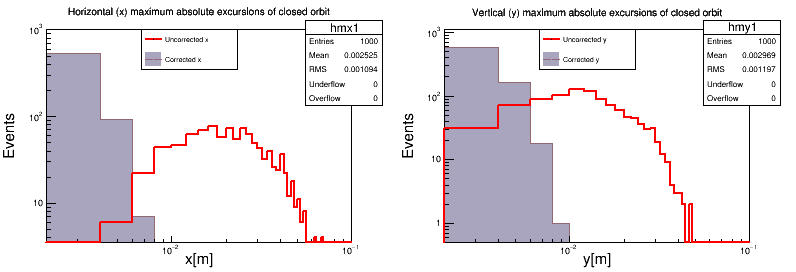}
\includegraphics[width=9.0cm]{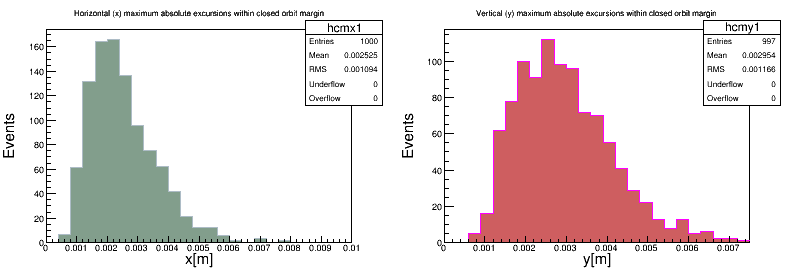}
\caption{(Color online) (Top) Maximum absolute excursions of the closed orbit before and after correction in both plane. 
(Bottom) Zoomed distributions of the excursions after correction within the global closed orbit margin, 7.5 mm, in horizontal 
and vertical planes respectively.}
\label{fig14}
\end{center}
\end{figure}
The requirement of closed orbit correction was set to be within 7.5 mm for both horizontal and vertical planes.
With a standard alignment technique, a 100.0~\% of the machines after correction in horizontal plane and a 99.7~\% 
in the vertical plane could be within the allowed global closed orbit margin, 7.5 mm, as shown in Fig.~\ref{fig14}. 
After correction, thus almost all the machines could be well within the stricter tolerances for the line. 
Fig.~\ref{fig12} shows the maximum absolute kick angle for both horizontal and vertical planes at highest carbon beam energy ( E = 430 MeV/u). 
As a result, the maximum required kick angle is foreseen to have the value less than 3.5 mrad in both plane. An average steering power
for the orbit correction of all static errors can be estimated as one root mean square (rms) and is less than 0.5 mrad. 
Accordingly the specification of correcting magnet is set to 5 mrad including margin as shown in Table~\ref{table:Spec}. 
\begin{figure}[h]
\begin{center}
\includegraphics[width=9.0cm]{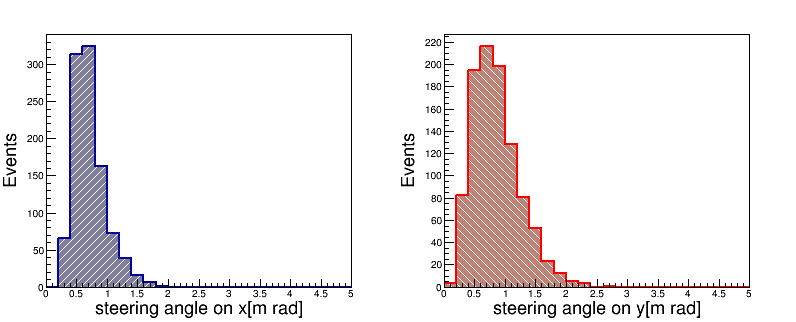}
\caption{(Color online) Maximum absolute steering angle for the horizontal and vertical planes at highest carbon beam energy, E = 430 MeV/u.}
\label{fig12}
\end{center}
\end{figure}

\begin{figure}
\begin{center}
\includegraphics[width=8.0cm]{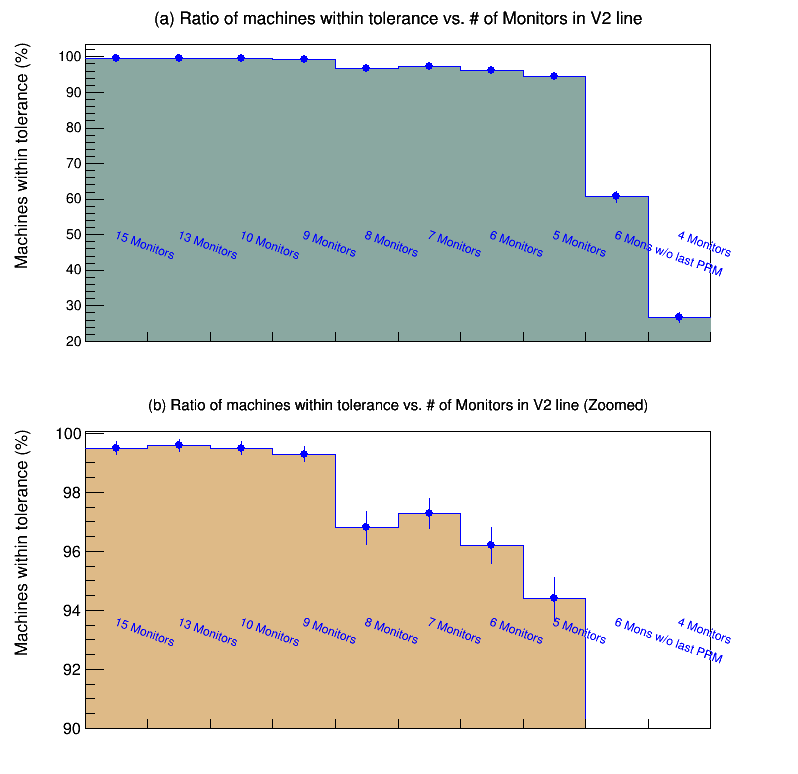}
\caption{(Color online) Ratio of the machines within tolerance with respect to the number of beam position monitor in the V2 line.}
\label{fig15}
\end{center}
\end{figure}
In order to optimize the alignment of corrector and beam position monitor, various combinations of corrector
and monitor are disposed with respect to the number of monitor as shown in Fig.~\ref{fig15}.
The optimized number of monitor in the beam line is determined to use 9 monitors.

\section{Conclusion}
The extracted transfer lines are based on the modular design taking into account the strong asymmetry 
between two transverse emittances and depend on the trapezoidal distribution of the beam in horizontal phase space. 
The KHIMA HEBT line has been fundamentally designed by using the WinAgile/Mad-X code for the whole beam lines up to each treatment room.
Based on the error analysis for every single line,
the optimized correction setup is settled and the specifications for the correcting magnets of the HEBT lines are determined.
A specification of corrector must fulfil the requirement for highest carbon beam energy,
430 MeV/u. Table~\ref{table:Spec} summarizes the specification of corrector used in the HEBT beam lines. 
\begin{table}
\caption{Specification for the correcting magnet of the HEBT line.}
\begin{ruledtabular}
\begin{tabular}{c|r|r}
Parameters                   &   Values &  Unit \\ \hline
Maximum rigidity (B$\rho$) & 6.62 & Tm\\
Max. deflection angle ($\theta$)& 5 & mrad \\
Effective length ($L_{eff}$)& 0.3 & m \\
Max. magnetic field  & 0.11 & T\\
( B = $\theta /L_{eff}$ $\times$ B$\rho$) & & \\
$\Delta$BL/BL & $< 2 \times 10^{-3}$ & \\
Field stability & 500 & ppm \\
No. of corrector & 23 & ea. \\
\end{tabular}
\end{ruledtabular}
\label{table:Spec}
\end{table}

\begin{acknowledgments}
This work was supported by the National Research and Development Program through the Korea Institute of Radiological and Medical Sciences funded by the Ministry of Science, ICT \& Future Planning (NRF-2015M2C3A1001637).
\end{acknowledgments}



\begin{references}
\bibitem{HJYim} Heejoong Yim {\it et al.}, J. Korean Phys. Soc. {\bf 67}, 1364(2015). 
\bibitem{Extract} M. Benedikt {\it et al.}, Nucl. Instrum. Methods Phys. Res. A {\bf 430}, 523 (1999). 
\bibitem{PIMMS} P. J. Bryant {\it et al.}, Proton Ion Medical Machine Study (PIMMS) Part 1 and 2, CERN/PS 1999-01-DI (1999) and CERN/PS 2000-007-
DR, Geneva (2000).
\bibitem{MBenedikt} M. Benedikt and A. Wrulich, ``MedAdustron-Project overview and status'', Eur. Phys. J. Plus {\bf 126: 69} (2011).
\bibitem{SRossi} S. Rossi {\it et al.}, ``The Status of CNAO'', Eur. Phys. J. Plus {\bf 126: 78} (2011).
\bibitem{MBenedikt01} M. Benedikt, ``Optics design of the extraction lines for the MedAustron hadron therapy centre'', 
Nucl. Instrum. Methods A 539 {\bf 25-36} (2004).
\bibitem{Badano} L. Badano {\it et al.}, "Closed orbit prognosis, correction and manipulation for the PIMMS synchrotron'', CERN/PS 99-059 (1999).
\bibitem{Agile} P. J. Bryant, AGILE program for synchrotron lattice design, http://nicewwww.cern.ch/~bryant.
\bibitem{Schmidt} F. Schmidt and H. Grote, ``MAD X an update from MAD8'', Proc. Part. Acc. Conference, Portland, U.S.A, 12.-16.5, {\bf 3497} (2003).
\bibitem{Herr} W. Herr, ``Implemenation of new closed orbit correction procedures in the MAD-X program'', CERN/SL 2002-48 (2002).
\bibitem{SVD} K. Baker, Singular Value Decomposition Tutorial, March 29, (2005).
\end{references}
\end{document}